\documentclass[aps,pra, twocolumn,groupaddress,twocolumn,10pt]{revtex4-1} 
\usepackage{amsmath}
\usepackage{latexsym}
\usepackage{graphicx}
\usepackage{amsfonts}
\usepackage{braket}
\usepackage{hyperref}
\usepackage{natbib}
\usepackage{amssymb}
\usepackage{nicefrac}
\usepackage{fancyhdr}
\usepackage[utf8]{inputenc}
\usepackage{dsfont}
\usepackage{colortbl}
\usepackage{xcolor}
\usepackage{subfigure}
\usepackage{psfrag}
\usepackage{tikz}
\usepackage{paralist}
\usepackage{nicefrac}
\usepackage{ulem}
\usetikzlibrary{arrows,shapes,decorations.pathmorphing}

\hypersetup{bookmarks=true, bookmarksopen=true, bookmarksopenlevel=1, unicode=false, pdftoolbar=true, pdfmenubar=true, pdffitwindow=false, pdfstartview={FitH}, pdftitle={Left-handed superlattice metamaterials for circuit-QED}, pdfauthor={Anette Messinger}, colorlinks=true, linkcolor=blue, citecolor=blue, urlcolor=blue}

\makeindex

\begin{document}

\title{Left-handed superlattice metamaterials for circuit QED}

\author{A. Messinger}
\affiliation{Theoretical Physics, Saarland University, 66123 Saarbr{\"u}cken, Germany}
\affiliation{School of Physics and Astronomy, University of Glasgow, Glasgow G12 8QQ, United Kingdom }
\author{B. G. Taketani}
\affiliation{Theoretical Physics, Saarland University, 66123 Saarbr{\"u}cken, Germany}
\affiliation{Departamento de F\'isica, Universidade Federal de Santa Catarina, 88040-900, Florian\'opolis, Brazil}
\author{F. K. Wilhelm}
\affiliation{Theoretical Physics, Saarland University, 66123 Saarbr{\"u}cken, Germany}

\newcommand{\w}{\omega}
\newcommand{\e}{\epsilon}

\newcommand{\SL}{\mathrm{sl}}
\newcommand{\IN}{\mathrm{in}}
\newcommand{\out}{\mathrm{out}}

\makeatletter
\def\Dated@name{Date: }
\makeatother
\date{\today}

\begin{abstract}
Quantum simulations is a promising field where a controllable system is used to mimic another system of interest, whose properties one wants to investigate. One of the key issues for such simulations is the ability to control the environment the system couples to, be it to isolate the system or to engineer a tailored environment of interest. One strategy recently put forward for environment engineering is the use of metamaterials with negative index of refraction. Here we build on this concept and propose a circuit-QED simulation of many-body Hamiltonians using superlattice metamaterials. We give a detailed description of a superlattice transmission line coupled to an embedded qubit, and show how this system can be used to simulate the spin-boson model in regimes where analytical and numerical methods usually fail, e.g. the strong coupling regime.
\end{abstract}

\maketitle

\section{Introduction}
One of the most promising applications of quantum technologies is the simulation of physical phenomena too complex to be dealt with by other techniques. Amongst the many possible physical implementations of such simulations, superconducting circuits take a central role, with experiments being able to engineer a large number of different many-body Hamiltonians \cite{Hur2016,Gu2017,Sameti2017,Lamata2017,Xu2018,Cosmic2018} and reaching regimes that are otherwise challenging for other platforms, e.g. the strong and ultrastrong coupling regime \cite{Wallraff2004,Lue2012,Han2016,Bosman2017,Braumueller2017,Moores2018,Leppakangas2018,Kockum2018}.

An important challenge for quantum simulations is the ability to engineer an adequate environment with which the simulated system interacts.  While some environments may have a simple, e.g. ohmic spectrum, some will have more complex, structured spectral densities. Reservoir engineering is a field that has found widespread applications, such as quantum state preparation \cite{Pielawa2010,Kienzler2014,Grimsmo2017}, steady-state entanglement generation \cite{Taketani2014,Vasco2016,Ockeloen-Korppi2018} and the study of light-matter interaction in structured photonic environment \cite{Liu2016,Haeberlein2015}. These usually rely on the creation of media with specific properties, control of the coupling of the system of interest to its environment or manipulation of the properties of existing environments. These approaches complement the use of lumped circuits to engineer decoherence \cite{vanderWal2003,Robertson2005}. One avenue recently suggested to devise media with a particular spectrum is the use of metamaterials, more specifically left-handed metamaterials \cite{Egger2013}. Contrary to regular, right-handed (RH) materials, the eigenfrequecies of left-handed (LH) materials increase with wavelengths \cite{Caloz2004}. Coupling RH and LH media one can thus find materials with new, interesting spectral properties. Such a hybrid material was shown to have a high density of modes at low frequencies \cite{Caloz2004a}, which could in turn be used to couple an embedded qubit to multiple environmental modes \cite{Egger2013}.

The ability to couple a system of interest in a controlled way, to an adjustable number of environmental modes opens the path to the quantum simulation of a myriad of different physical phenomena. In this paper we build on these ideas and show how left-handed superconducting superlattices can be used to investigate the phase diagram of the Spin-Boson model \cite{Caldeira:1981} with a novel structured environment. The superlattice structure investigated here (for which \cite{Egger2013} is a special case) leads to a 2-band spectrum, with the number of modes in each band given by the array length. This is used to control the number of modes with which an embedded qubit interacts, allowing for great flexibility on the design of the qubit's environment. As a testbed for our system, we present a detailed investigation of the phase structure of the spin-boson model and discover a rich phase diagram. Our results pinpoint LH superlattice metamaterials as a tool with interesting properties for microwave photonics.

This article is organized as follow: in sec. II we describe the coupled transmission line and determine its spectral properties. In sec III we investigate the interaction of the photonic modes with an embedded qubit. Our results are obtained by both an analytical and a numerical approach and we show the phase diagram under weak coupling. We present our concluding remarks in sec. IV.

\section{The System}
\begin{figure}
\includegraphics[width=1\columnwidth]{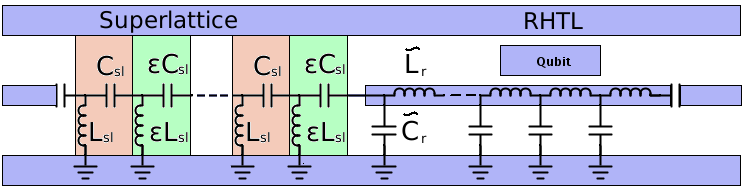}
\caption{\label{line}(Color online) Composite left-handed superlattice and right-handed transmission line coupled to a qubit. The superlattice unit cell is formed by two cells, indicated by red (dark gray) and green (light gray) background. The equivalent discrete circuit model for the RHTL is illustrated.}
\end{figure}

In this work we will investigate a superlattice structure consisting of LC resonators in series and its interaction with the system of interest, a superconducting qubit. A standard circuit transmission line (TL), e.g. a coaxial cable, can be modeled by an LC array of series inductors and grounded capacitors as shown in Fig.~\ref{line} (right side). In the continuum limit, where the LC unit-cell size tends to zero while the inductance and capacitance per unit length are kept constant, this array presents the dispersion of the TEM-modes of the transmission line \cite{Pozar2012}. For such a system, the group and phase velocities are oriented parallel and both energy and wavefronts travel away from the source. These are termed right-handed transmission lines (RHTL), as these parallel velocities stem from the electric and magnetic field vectors and the wave vector forming a right-handed set in three-dimensions \cite{Solymar:2014}. If we now invert these unit-cells, connecting the inductors to ground in parallel, while placing the capacitors in series, the resulting waves will have anti-parallel group and phase velocities. Now the electric and magnetic field vectors and the wave vector form a left-handed set \cite{Eleftheriades2002} and such metamaterials are called left-handed transmission lines (LHTL). The properties of LHTL include opposite group and phase velocities as well as a falling dispersion relation \cite{Caloz2004,Caloz2004a} and applications of left-handed materials range from cloaking \cite{Zheng2009,Yang2016} to a perfect lens \cite{Zharov2005,Rosenblatt2016}.

Recently it was shown that a composite transmission line, with left-handed and right-handed elements could be used to engineer the electromagnetic environment experienced by a superconducting qubit \cite{Egger2013}. It constructs on the fact that a pure left-handed transmission line shows a cutoff infrared frequency. Close to that cutoff frequency, the LHTL and therefore also the coupled transmission line has a high mode density. The RHTL has a linear dispersion relation an therefore does not support as many different modes as the LHTL close to the cutoff frequency. Therefore, all these modes share similar voltage profiles with only small variation of wavelengths when entering the RH part of the transmission line. This permits multimode strong coupling of a qubit embedded in the RH part of the line to the bosonic modes. The study proposed this system as a test bed to simulate the spin-boson model \cite{Leggett1987}.

In this paper we build on these ideas and consider the next step on this approach, namely a superlattice LHTL consisting of two alternating left-handed LC cells with different frequencies $\frac{1}{\sqrt{LC}}$, where $L=L_\SL$ ($L^\prime_\SL$) and $C=C_\SL$ ($C^\prime_\SL$) are the inductance and capacitance of the first (second) cell  (see Fig. \ref{line}). To avoid unwanted reflections between cells the characteristic impedance $Z=\sqrt{\frac{L}{C}} $ must match, $Z_\SL=Z^\prime_\SL$ \cite{Pozar2012}. This means the inductance must change between the two cells in the same way as the capacitance. We introduce the parameter $\epsilon\in\mathbb{R}$ to quantify this ratio and set  $ L^\prime_\SL=\epsilon L_\SL $ and $ C^\prime_\SL=\epsilon C_\SL$. This superlattice LHTL will then couple directly to a RH transmission line, impedance matched to the characteristic impedance of the superlattice. In contrast to the metamaterial LHTL which has to be created with discrete circuit elements, the right-handed part can be a simple coplanar waveguide.

The above composite transmission line is coupled to the (high-temperature) control and measuring part of the experiment trough coupling capacitors. Finally, a superconducting qubit inside the RHTL can be designed to couple to the bosonic modes. We shall now describe properties of the composite transmission line and in the next section give details on the qubit and its effective dynamics.

\subsection{Dispersion relation}
\label{sec:disp_rel}
For the discrete RHTL, the dispersion relation can be found to be
\begin{gather}
\w(k_{r})=\frac{2}{\sqrt{\tilde{L}_{\rm r}\tilde{C}_{\rm r}}}\sin\left( \frac{k_{\rm r}d_{{\rm r}}}{2n_{\rm r}}\right),
\label{eq:disp_r}\\
k_{\rm r,max}=\pi \frac{n_{\rm r}}{d_{\rm r}}
\end{gather}
where $ \tilde{L}_{\rm r} $ and $ \tilde{C}_{\rm r} $ are the corresponding cell inductance and capacitance, $ d_{\rm r} $ is the total length of the line and $ n_{\rm r} $ is the number of cells. Taking the limit $ n_{\rm r}\rightarrow\infty $ while keeping the inductance and capacitance per unit length constant, equation (\ref{eq:disp_r}) reduces to the usual linear dispersion
\begin{equation}
\w(k_{\rm r})\approx\frac{k_{\rm r}d_{\rm r}}{n_{\rm r}\sqrt{\tilde{C}_{\rm r}\tilde{L}_{\rm r}}}=\frac{k_{\rm r}}{\sqrt{c_{\rm r}l_{\rm r}}},
\label{eq:disp_r_cont}
\end{equation}
with $ c_{\rm r} $ and $ l_{\rm r} $ being capacitance and inductance per unit length.

The dispersion relation of the superlattice can be obtained via its ABCD-matrix, $\mathbf{b}$, which connects incoming and outgoing voltages and currents of circuit elements as shown in Fig. \ref{VI} \cite{Solymar:2014}:
\begin{figure}
\includegraphics[width=0.7\columnwidth]{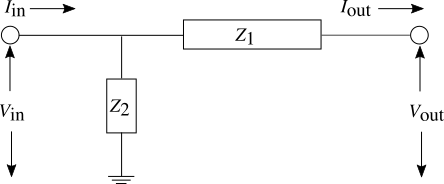}
\caption{\label{VI} Incoming and outgoing voltage and current of lattice cell with impedances $ Z_1 $ and $ Z_2 $. The matrix describing the relations between outgoing and incoming values in terms of $ Z_1 $ and $ Z_2 $ is the ABCD-matrix.}
\end{figure}
\begin{equation}
\begin{pmatrix}V_{\rm out}\\
I_{\rm out}
\end{pmatrix}=\mathbf{b}\cdot\begin{pmatrix}V_{\rm in}\\
I_{\rm in}
\end{pmatrix},
\label{eq:ABCDallgem} 
\end{equation}
where $V_\IN$ and $I_\IN$ ($V_\out$ and $I_\out$) are respectively the input (output) voltage and current. For an arbitrary circuit consisting of a series element with impedance $ Z_1$ connected in parallel to an element with impedance $ Z_2$ shunted to ground, currents and voltages are related as
\begin{align*}
Z_1&=\frac{V_1}{I_1}=\frac{V_{\rm out}-V_{\rm in}}{I_{\rm in}}\\
Z_2&=\frac{V_2}{I_2}=\frac{V_{\rm in}}{I_{\rm out}-I_{\rm in}},
\end{align*}
where the sub-indices $1$ and $2$ denote current or voltage across the corresponding element. Since our lattice cells consist of a series capacitance and an inductance to ground, we can write
\begin{align*}
Z_1=\frac{1}{i\w C_{\rm sl}},\;\; Z_2=i\w L_{\rm sl},
\end{align*}
for the first lattice cell. The second cell is found by substituting the corresponding capacitance and inductance. The resulting ABCD-matrices are
\begin{equation*}
\mathbf{b_{A}}=\left(\begin{matrix}1-\frac{\omega_{\rm sl}^{2}}{\w^{2}} & \frac{1}{i\w C_{{\rm sl}}}\\
\\
\frac{1}{i\w L_{{\rm sl}}} & 1
\end{matrix}\right),\,\,\mathbf{b_{B}}=\left(\begin{matrix}1-\frac{\omega_{\rm sl}^{2}}{\w^{2}\e^{2}} & \frac{1}{i\w\e C_{{\rm sl}}}\\
\\
\frac{1}{i\w\e L_{{\rm sl}}} & 1
\end{matrix}\right) ,
\end{equation*}
where $ \w_{\rm sl}=\frac{1}{\sqrt{L_{\rm sl}C_{\rm sl}}} $ is the resonance frequency of the first cell. The matrix of a supercell, the most fundamental building block of the superlattice, is given by the product of the two single cell matrices
\begin{align}
\mathbf{b} &= \mathbf{b_{A}b_{B}} \notag \\
&= \left(\begin{matrix}1-\frac{\omega_{\rm sl}^{2}}{\w^{2}}(1+\frac{1}{\e}+\frac{1}{\e^{2}}-\frac{\omega_{\rm sl}^{2}}{\w^{2}\e^{2}}) & \frac{1}{i\w C_{{\rm sl}}}(1+\frac{1}{\e}-\frac{\omega_{\rm sl}^{2}}{\w^{2}\e})\\
\\
\frac{1}{i\w L_{{\rm sl}}}(1+\frac{1}{\e}-\frac{\omega_{\rm sl}^{2}}{\w^{2}\e^{2}}) & 1-\frac{\omega_{\rm sl}^{2}}{\w^{2}\e}
\end{matrix}\right).\label{eq:ABCD} 
\end{align}

To find the dispersion relation for the superlattice array we use a plane wave ansatz
\begin{align}
V(z,t)&=V_{0}e^{i(kz-\w t)} 
\\I(z,t)&=I_{0}e^{i(kz-\w t)}.
\end{align}
For a supercell of size $\Delta z$, we see that, at fixed times
\begin{align}
V_{\rm out}&=e^{-ik\Delta z}V_{\rm in}\label{eq:phaseV} \\
I_{\rm out}&=e^{-ik\Delta z}I_{\rm in}.
\label{eq:phaseI} 
\end{align}
Comparing equations (\ref{eq:phaseV}) and (\ref{eq:phaseI}) with the ABCD-matrix (\ref{eq:ABCDallgem}), we obtain
\begin{align*}
e^{-ik\Delta z}V_{in}&=b_{11}V_{in}+b_{12}I_{in}\\e^{-ik\Delta z}I_{in}&=b_{21}V_{in}+b_{22}I_{in}.
\end{align*}
Here $b_{ij}$ are the $\{i,j\}$ elements of matrix $\mathbf b$. This readily leads to
\begin{align}
(e^{-ik\Delta z}-b_{11})(e^{-ik\Delta z}-b_{22})=b_{12}b_{21}.
\label{eq:dispSchritt}
\end{align}
All elements in the superlattice circuit behave independent of the direction in which current flows, therefore, from reciprocity arguments, it follows that $ b_{11}b_{22}-b_{12}b_{21}=1 $ must be fulfilled \cite{Solymar:2014}. Equation (\ref{eq:dispSchritt}) now simplifies to
\begin{equation}
b_{11}+b_{22}=2\cos(k\Delta z),
\end{equation}
from which the dispersion relation is found to be
\begin{equation}
\w(k_{\rm sl})=\sqrt{\frac{\omega_{\rm sl}^{2}}{\frac{(1+\e)^{2}}{2}\pm\sqrt{\epsilon^{2}(2\cos(k_{\rm sl}\Delta z)-2)+\frac{(1+\e)^{4}}{4}}}}.
\label{eq:disp_sl}
\end{equation}
Appendix \ref{app:EL_formalism} describes an alternative formalism to obtain the dispersion relation based on Euler-Lagrange equations. 

\begin{figure}[th]
	\centering
	\includegraphics[width=\columnwidth]{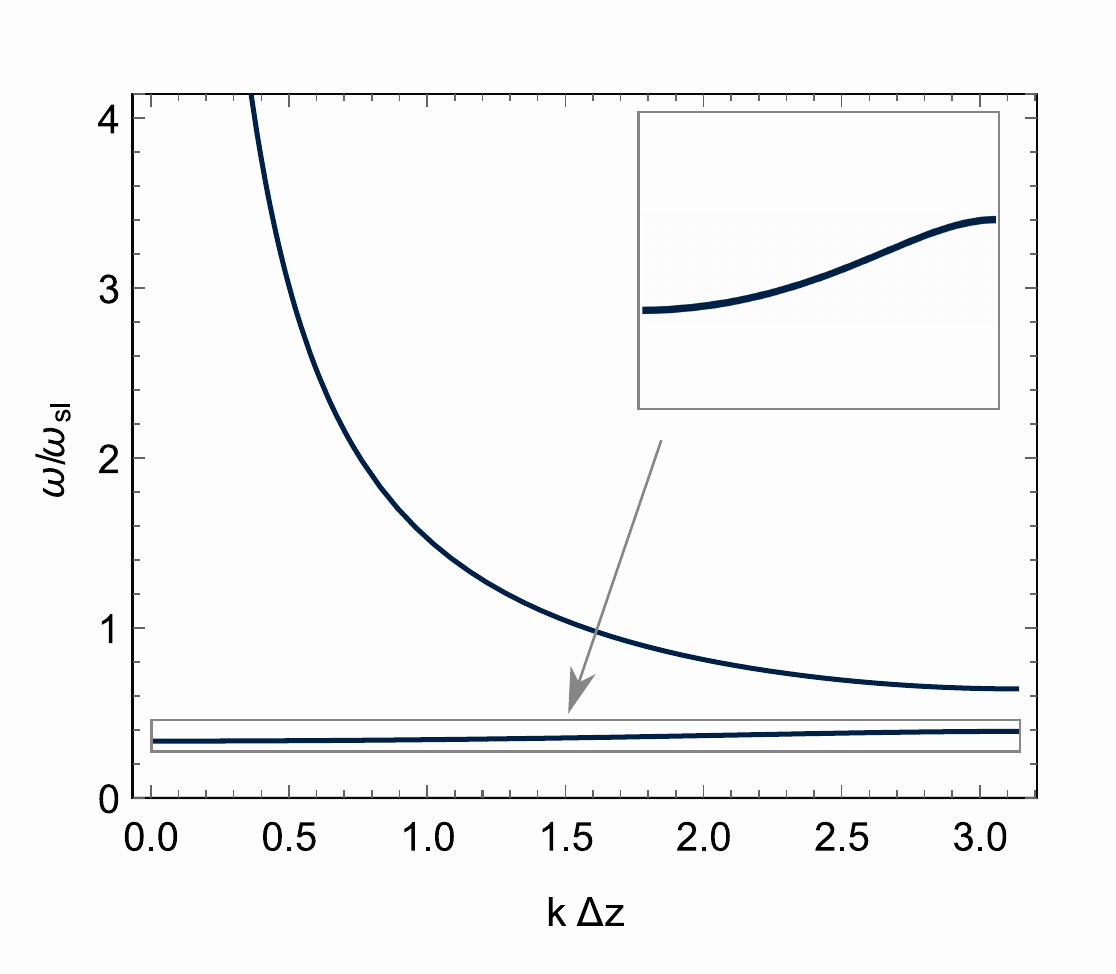}
	\caption{ Mode number vs. frequency of a left-handed superlattice transmission line with $\epsilon=2$. The periodic structure gives rise to a band gap with the high frequency band revealing the left-handedness of the system.}
	\label{fig:disp_rel}
\end{figure}

\begin{figure}[th]
	\centering
	\includegraphics[width=\columnwidth]{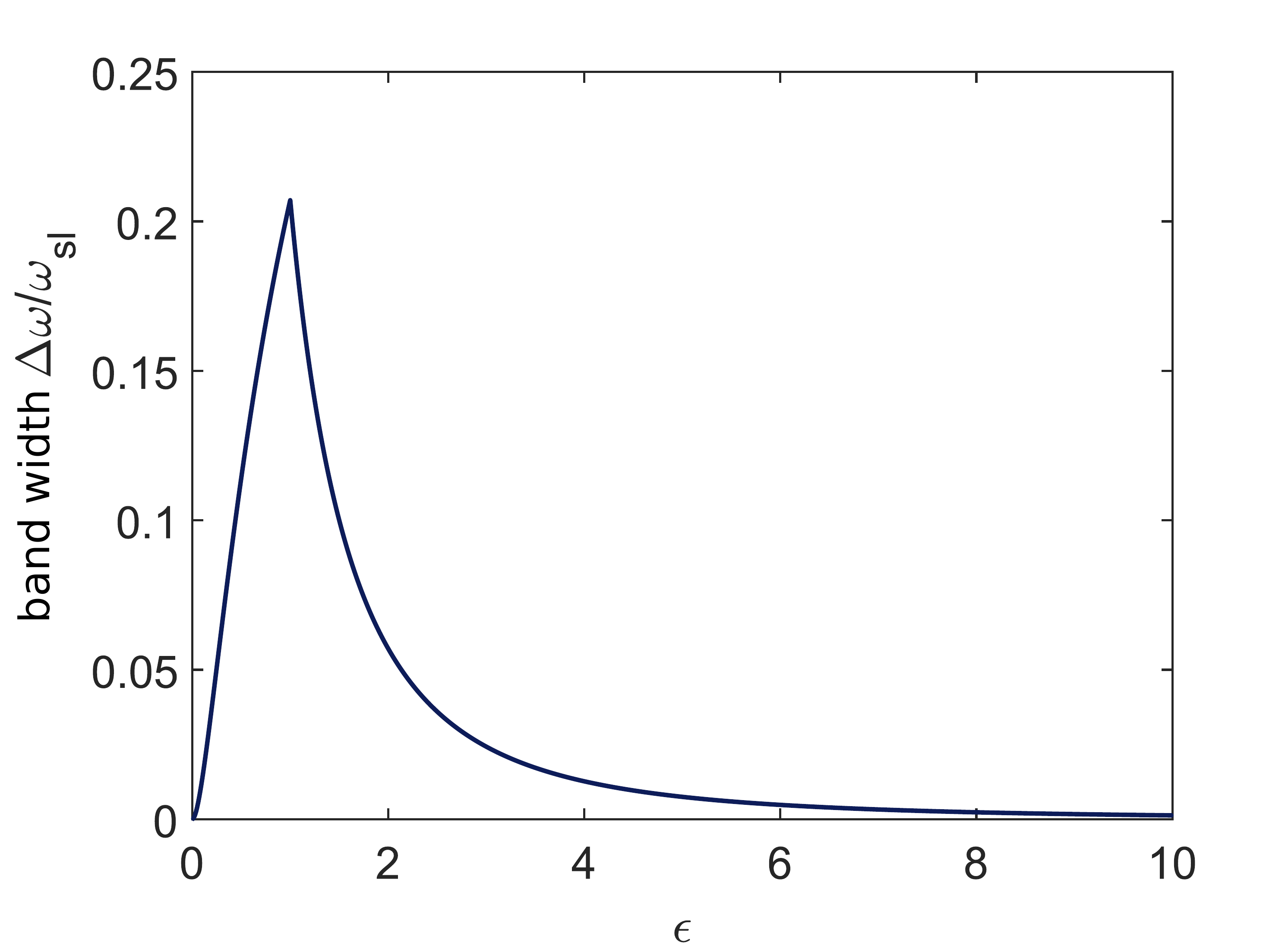}
	\caption{Width of the lower frequency band as a function of the superlattice parameter $\epsilon$.}
	\label{fig:band_width}
\end{figure}

Naturally, as shown in Fig.~\ref{fig:disp_rel}, the superlattice gives rise to a frequency band gap, which can be found by looking at the codomain of the cosine in the superlattice dispersion relation. The two bands are limited to 
\begin{equation*}
 \omega(k) \in [\omega_{\rm 1-},\omega_{\rm 1+}]\,\cup\,[\omega_2,\infty)
 \end{equation*}
 with 
 \begin{align*}
& \omega_{\rm 1-}=\frac{\omega_{\rm sl}}{1+\epsilon},\\
 &\omega_{\rm 1+}=\frac{\omega_{\rm sl}}{\sqrt{\frac{1}{2}(1+\epsilon)^{2}+\sqrt{\frac{1}{4}(1+\epsilon)^{4}-4\epsilon^{2}}}},\\
 &\omega_2=\frac{\w_{sl}}{\sqrt{\frac{1}{2}(1+\e)^{2}-\sqrt{\frac{1}{4}(1+\e)^{4}-4\e^{2}}}}. 
 \end{align*}
This band gap can also be understood as a result of destructive interference from Bragg reflection due to the two different cells the superlattice consists of. Similar effects are known from photonic crystals or phonons in diatomic lattices \cite{Joannopoulos2008}. Moreover, a signature of the left-handedness of the coupled line is present in the upper band, where the frequency decays with growing wave number.
Figure \ref{fig:band_width} shows the dependency of the bandwidth $\Delta\omega=\omega_{\rm 1+}-\omega_{\rm 1-}$ on the ratio $\epsilon$ of the two superlattice resonance frequencies. One sees that already moderate changes of $\epsilon$ are enough to decrease the bandwidth and therefore increase the mode density by orders of magnitude.

\subsection{Eigenmodes of the hybrid transmission line}
Equations (\ref{eq:disp_r_cont}) and (\ref{eq:disp_sl}) relate the wave number and frequency for a plane wave solution inside the RH and LHTLs. Now let us see what happens when we couple both to create a hybrid transmission line. In order to find the desired eigenmodes, we use a plane wave ansatz with waves traveling in both directions, because we now consider a system of finite length and reflections at the input and output ports. The currents and voltages in each part of the transmission line are thus
\begin{align*}
I_{{\rm sl}}(z,t)&=I_{0}(e^{ik_{\rm sl}z}+\alpha_{1}e^{-ik_{\rm sl}z})e^{-i\w t}\\
I_{\rm r}(z,t)&=I_{0}(\alpha_{2}e^{ik_{\rm r}z}+\alpha_{3}e^{-ik_{\rm r}z})e^{-i\w t}\\
V_{{\rm sl}}(z,t)&=V_{0}(e^{ik_{\rm sl}z}+\beta_{1}e^{-ik_{\rm sl}z})e^{-i\w t}\\
V_{{\rm r}}(z,t)&=V_{0}(\beta_{2}e^{ik_{\rm r}z}+\beta_{3}e^{-ik_{\rm r}z})e^{-i\w t}.
\end{align*}
Naturally, the frequency $\w$ must be the same in the RHTL and the superlattice to fulfil energy conservation, whereas the left- and right-handed wave numbers $k_{\rm sl}$ and $k_{\rm r}$ may differ, by virtue of the different index of refraction of the two separate TLs. 

The unknown coefficients are found from the boundary conditions $ I_{{\rm sl}}(0,t)=I_{\rm r}(d_{{\rm sl}}+d_{{\rm r}},t)=0 $ (currents with nodes at the input and output ports) and $ I_{{\rm sl}}(d_{{\rm sl}},t)=I_{\rm r}(d_{{\rm sl}},t) $ and $ V_{{\rm sl}}(d_{{\rm sl}},t)=V_{{\rm r}}(d_{{\rm sl}},t) $ (continuity of current and voltage at the coupling between the lines). In addition, in the RHTL we use the characteristic impedance $ Z_{r}=\sqrt{\frac{L_{{\rm r}}}{C_{{\rm r}}}}=\frac{V^{+}}{I^{+}}=-\frac{V^{-}}{I^{-}}$, which relates the amplitudes $ V^{\pm}=V_{0}\beta_{2/3} $ of right, $(+)$, and left, $(-)$, travelling waves \cite{Pozar2012}. In the superlattice, we use the ABCD-matrix to relate voltage and current before and after a supercell, similarly to section \ref{sec:disp_rel}. This leads to an over-determined system of equations for $ \alpha_i,\, \beta_i $ and $ V_0 $ ($ I_0 $ is chosen as a free parameter). Leaving out the voltage coupling condition, we find the following wave equations:
\begin{align*}
I_{{\rm sl}}(z,t)&=I_{0}(e^{ik_{\rm sl}z}-e^{-ik_{\rm sl}z})e^{-i\w t}\notag\\
I_{\rm r}(z,t)&=I_{0}\alpha(e^{-ik_{\rm r}d}e^{ik_{\rm r}z}-e^{ik_{\rm r}d}e^{-ik_{\rm r}z})e^{-i\w t}\notag\\
V_{\rm sl}(z,t)&=Z_{\rm sl}I_{0}(e^{ik_{\rm sl}z}+\beta e^{-ik_{\rm sl}z})e^{-i\w t}\notag\\
V_{\rm r}(z,t)&=Z_{\rm r}I_{0}\alpha(e^{-ik_{\rm r}d}e^{ik_{\rm r}z}+e^{ik_{\rm r}d}e^{-ik_{\rm r}z})e^{-i\w t},
\end{align*}
with $ \alpha=-\frac{\sin(k_{\rm sl}d_{{\rm sl}})}{\sin(k_{\rm r}d_{{\rm r}})} $, $ \beta=-\frac{e^{-ik_{\rm sl}\Delta z}-b_{22}}{e^{ik_{\rm sl}\Delta z}-b_{22}} $, $ Z_{\rm sl}=\frac{e^{ik_{\rm sl}\Delta z}-b_{22}}{b_{21}}$ and $ d=d_{{\rm r}}+d_{{\rm sl}} $. From the last condition, we finally find the self-consistency equation 
\begin{equation*}
2Z_{\rm r}\alpha\cos(k_{\rm r}d_{{\rm r}})-Z_{\rm sl}(e^{ik_{\rm sl}d_{{\rm sl}}}+\beta e^{-ik_{\rm sl}d_{{\rm sl}}})\overset{!}{=}0,
\end{equation*}
which can be solved numerically for the frequency $\w$ by using equations (\ref{eq:disp_r_cont}) and (\ref{eq:disp_sl}).

\subsection{Density of Modes}
\label{sec:dom}
The high density of modes (DOM) observed in the lower band of the superlattice spectrum, Fig.~\ref{fig:disp_rel}, is the main ingredient required for the type of many-body quantum simulation we wish to study. However, evaluation of the DOM of the hybrid line is made difficult due to its discrete nature. In the following we use both a numerical method and an analytical approximation to find the DOM. All numerical calculations in this work are performed with $\omega_{\rm sl}=1/\sqrt{2\cdot6\cdot10^{-23}}\rm{Hz}\approx91.3\rm{GHz}$, $\omega_{\rm r}=1/\sqrt{2.5\cdot7.5\cdot10^{-23}}\rm{Hz}\approx73.0\rm{GHz}$ and $Z=\sqrt{3\cdot10^{3}}\Omega\approx54.8\Omega$. For the numerical approximation, as the modes are equally spaced in $k$-space we can set the DOM at a given frequency $\w_i$ by the frequency difference to its neighbouring modes 
\begin{equation}
D_{\rm num}(\w_{i})=\frac{2}{\w_{i+1}-\w_{i-1}},
\end{equation}
for $i>1$. We thus set the density at frequency $\w_i$ to the average density in the region between $\w_{i-1}$ and $\w_{i+1}$, which contains exactly two modes. It is important to mention that this calculation will fail not only at $i=0$ but also for the modes at the band edges, where the next or previous mode is across the band gap. Therefore we shall not consider these modes when determining the density of modes by this method.

An analytical expression can be found by calculating the individual DOM for the decoupled RH, $D_{\rm r}(\w)$, and superlattice, $D_{\SL}(\w)$, transmission lines and approximating the DOM for the coupled system by $D(\w)\approx D_{\rm r}(\w)+D_{\SL}(\w)$. For the decoupled transmission lines, the DOM can be calculated from
\begin{equation*}
 D(\w)=\frac{{\rm d}n}{{\rm d}k}\frac{{\rm d}k}{{\rm d}\w},
\end{equation*}
using the inverse of the dispersion relations (\ref{eq:disp_r_cont}) and (\ref{eq:disp_sl}) and the fact that wave vectors in the decoupled systems are equally spaced with $k_{\rm r/sl}=\frac{n\pi}{l_{\rm r/sl}}$ and $n\in\mathbb{N}$. The derivatives of the inverse dispersion relations are
\begin{equation*}
 \frac{{\rm d}k_{r}}{{\rm d}\w}=\frac{1}{\w_{\rm r}l_{{\rm r}}},
\end{equation*}
\begin{equation*}
\frac{{\rm d}k_{\rm sl}}{{\rm d}\w}=\frac{\frac{\w_{\rm sl}^{2}}{\w^{3}}\left(1+\frac{1}{\e}\right)^{2}-\frac{\w_{\rm sl}^{4}}{\w^{5}}\frac{2}{\e^{2}}}{\Delta z\sqrt{1-\frac{1}{4}\left(2+\frac{\w_{\rm sl}^{4}}{\w^{4}}\frac{1}{\e^{2}}-\frac{\w_{\rm sl}^{2}}{\w^{2}}\left(1+\frac{1}{\e}\right)^{2}\right)^{2}}}.
\end{equation*}
With these we obtain the approximate DOM of the coupled system $ D(\w)=\frac{l_{\rm r}}{\pi} \frac{{\rm d}k_{r}}{{\rm d}\w}+\frac{l_{\rm sl}}{\pi} \frac{{\rm d}k_{\rm sl}}{{\rm d}\w} $, with $ \frac{l_{\rm sl}}{\Delta z}=n_{\rm sl} $ being the number of supercells in the superlattice.

\begin{figure}[th]
\centering
\includegraphics[width=\columnwidth]{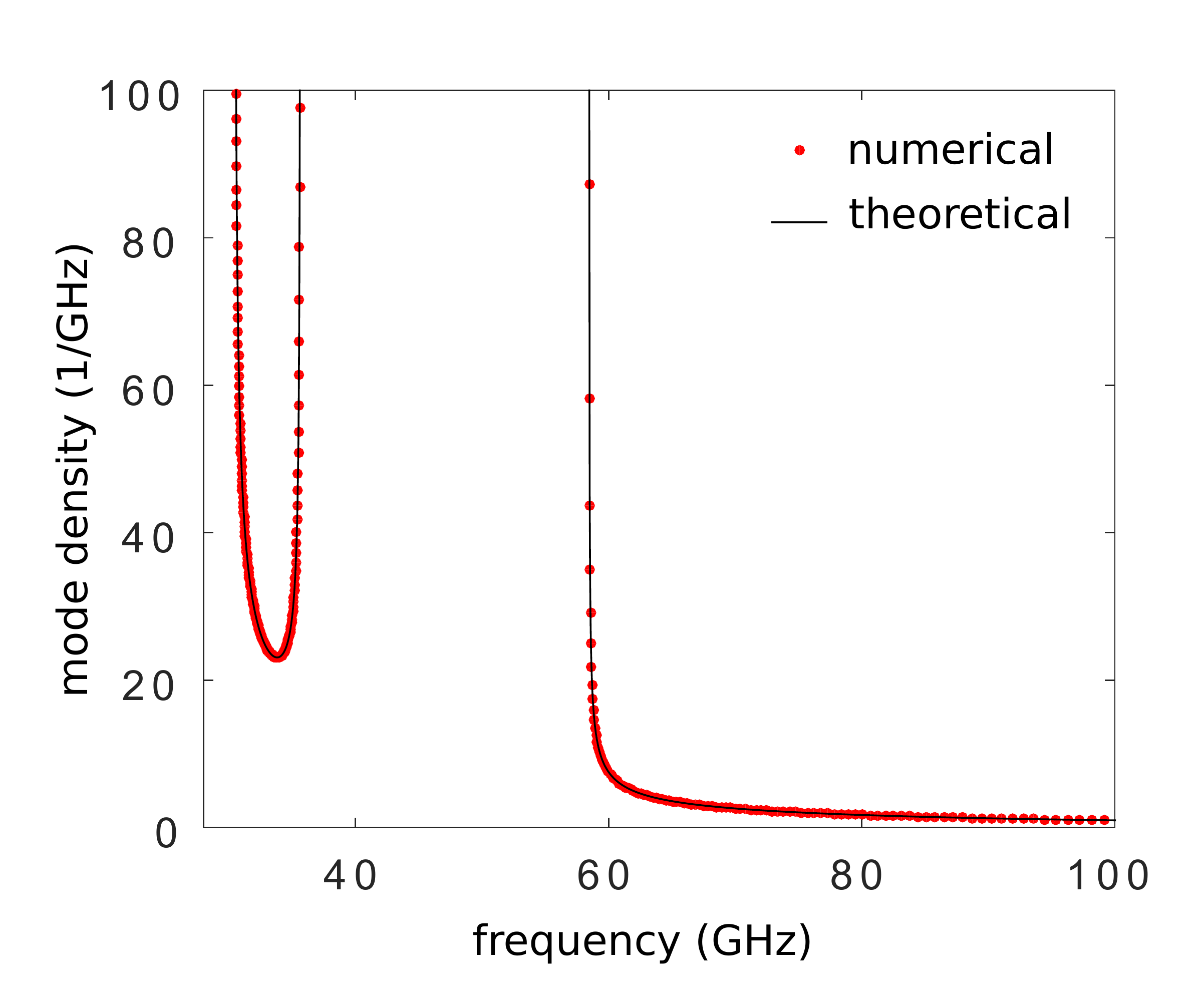}
\caption{(Color online) Numerical (red dots) and theoretical (black line) approximations to the mode density $ D(\w)$ of the coupled system for a superlattice with 200 cells and $\epsilon=2$.}
\label{fig:dom200}
\end{figure}

Figure~\ref{fig:dom200} shows the numerical and analytical approximation for the DOM with $n_{\rm sl}=200$ supercells. The agreement between these two independent approaches gives strong indication that both are valid approximations. The increased DOM in the low frequency region reflects the small width of the superlattice energy band and can easily be increased further:  As the number of modes in each band is independent of the band width, narrowing the band implies a higher density of modes. As shown in Fig.~\ref{fig:band_width}, the width of the low frequency band has a maximum when the cells forming the supercell are identical (the array is not a superlattice) and the band narrows as the difference between the frequencies of these cells increases.The Van Hove singularities will play an important role in the phase diagram studied in Section \ref{subsec:PhaseDiag}.

\subsection{Voltage Profile} 
\label{sec:voltage}
The high mode density observed above will be used in the next section to allow a qubit to couple to multiple cavity modes simultaneously. The qubit-cavity interaction depends not only on their detuning, but also on their natural coupling strength. For a flux qubit it will depend on each mode's current profile, which we will now analyse. For a qubit embedded in a standard cavity, the mode density at high frequencies can be increased simply by making the cavity longer \cite{Sundaresan2015}, thus allowing the qubit to couple to many modes. However, if the qubit is positioned so as to maximize its coupling to a given low frequency mode, the coupling to the following mode will have a stark decrease due to the different voltage profile along the right-handed line.

\begin{figure}[th]
\centering
\includegraphics[width=\columnwidth]{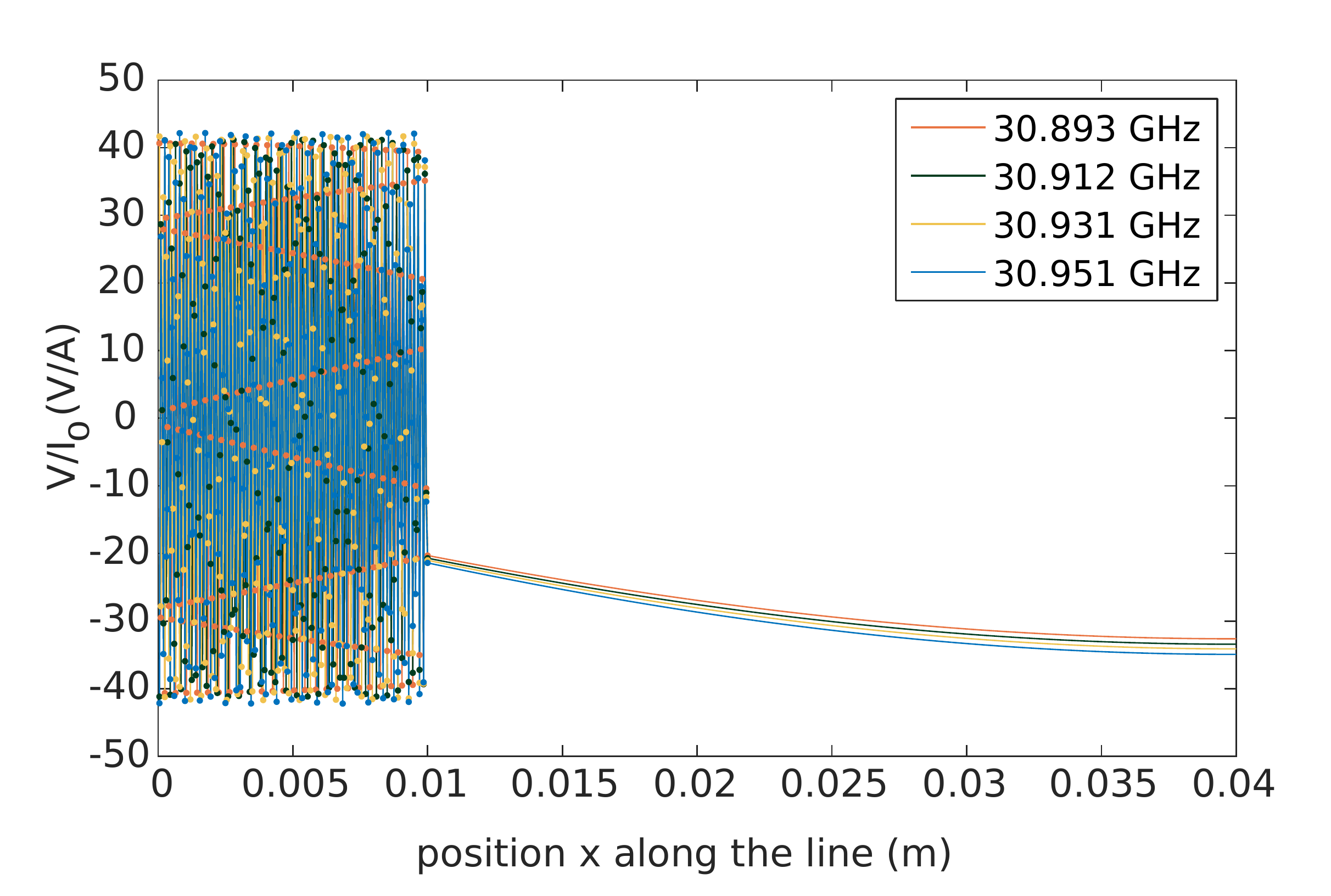}
\caption{(Color online) Voltage profile for the modes with index 50 to 53 for a transmission line with $ n_{\rm sl}=200 $ and $ \e=2 $. The superlattice is positioned at the left side and the RHTL on the right side in the plot with their coupling at $ 0.01m $. Due to the discreteness of the superlattice, the lines do not represent the real voltage profile for each position along the circuit but only at circuit nodes which are represented by dots.}
\label{fig:profile}
\end{figure}

The hybrid transmission line considered here gives rise to an altogether different behaviour. First, the high density modes appear at low frequency. Second, as shown in Fig.\ref{fig:profile}, low frequency neighbouring modes show sharp differences in their voltage/current profiles inside the superlattice, but remarkably similar profiles within the right-handed medium. This feature suggests that neighbouring modes can be made to simultaneously have comparable coupling strengths and detuning to the qubit. Put together, these tools allow us great versatility to engineer distinct qubit environments.

\section{Spin-Boson model}
To illustrate the use of the composite transmission line, we will now use it to investigate the ground state of the qubit interacting with the bosonic multimode environment characterized above. The composite system is described by the Hamiltonian 
\begin{equation}
\hat{H}=\hbar\left(-\frac{\Delta_{0}}{2}\hat{\sigma}_{x}+\underset{k}{\sum}\omega_{k}\hat{a}_{k}^{\dagger}\hat{a}_{k}+\hat{\sigma}_{z}\underset{k}{\sum}g(\hat{a}_{k}^{\dagger}+\hat{a}_{k})\right),
\label{eq:hamilton}
\end{equation}
where the first two terms describe the qubit and environment free Hamiltonians and the last term represents their coupling. Here the qubit is taken to be degenerate and $\Delta_0$ is the tunnelling rate between the bare qubit energy eigenstates $\ket L$ and $\ket R$. This Hamiltonian describes the spin-boson model (SBM), which is a standard model for dissipative two-level systems~\cite{Caldeira:1981}. In the usual SBM the energy eigenstates represent the position of a particle in a double-well potential, i.e., $\ket L$ ($\ket R$) being a qubit in the left (right) quantum well. This model admits two qubit quantum phases, a localized phase ($\Delta_0=0$), with the qubit having no intrinsic dynamics and localizing in one of the two wells, and a delocalized phase, when $\Delta_0\neq0$ and the system displaying Rabi oscillations between the two wells~\cite{Florens2010}. As discussed above, the low frequency mode profiles are very similar close to the qubit's position. Therefore we will take qubit to have equal coupling constant to all modes.

\subsection{Adiabatic renormalization}
\label{sec:ad_renorm} 
The qubit multimode coupling renders Hamiltonian (\ref{eq:hamilton}) difficult to diagonalize. However, as some modes have frequencies much larger than that of the qubit, they adiabatically follow the qubit's dynamics. Adiabatic renormalization techniques can thus be used to transform the Hamiltonian and find an effective tunneling element.

Let us first look at the $\Delta_{0}=0$ case. This Hamiltonian can be fully diagonalized by the unitary transformation
\begin{equation}
\hat{U}=\exp\left(-\hat{\sigma}_{z}\underset{k}{\sum}\frac{g}{\w_{k}}(\hat{a}_{k}-\hat{a}_{k}^{\dagger})\right),
\label{fulltrafo}
\end{equation}
which leads to
\begin{equation*}
\hat{U}\hat{H}(\Delta_{0}=0)\hat{U}^{\dagger}=\underset{k}{\sum}\omega_{k}\hat{a}_{k}^{\dagger}\hat{a}_{k}-\underset{k}{\sum}\frac{g^{2}}{\w_{k}}.
\end{equation*}
The adiabatic renormalization procedure consists of iterative steps, where at each step we adiabatically eliminate modes whose frequencies are higher than any qubit frequency, i.e., modes that adiabatically follow the qubit's dynamics. To begin, we transform Hamiltonian (\ref{eq:hamilton}) using
\begin{equation}
\hat{U}_{1}=\exp\left(-\hat{\sigma}_{z}\underset{\omega_k>\Delta_{0}}{\sum}\frac{g}{\w_{k}}(\hat{a}_{k}-\hat{a}_{k}^{\dagger})\right),
\label{trafostep}
\end{equation}
where the summation is taken over high frequency modes, $\w_k>\Delta_0$. By rewriting this operator as
\begin{align*}
\hat{U}_1&=\cos\left(i\underset{\omega_k>\Delta_{0}}{\sum}\frac{g}{\w_{k}}(\hat{a}_{k}-\hat{a}_{k}^{\dagger})\right)\\
&+i\hat{\sigma}_{z}\sin\left(i\underset{\omega_k>\Delta_{0}}{\sum}\frac{g}{\w_{k}}(\hat{a}_{k}-\hat{a}_{k}^{\dagger})\right),
\end{align*}
we find that $ \hat{\sigma}_x $ transforms as
\begin{align*}
\hat{U}_{1}\hat{\sigma}_{x}U_{1}^{\dagger}&=\cos\left(i\underset{\omega_k>\Delta_{0}}{\sum}\frac{g}{\w_{k}}(\hat{a}_{k}-\hat{a}_{k}^{\dagger})\right)\hat{\sigma}_{x}\\
&+\sin\left(i\underset{\omega_k>\Delta_{0}}{\sum}\frac{g}{\w_{k}}(\hat{a}_{k}-\hat{a}_{k}^{\dagger})\right)\hat{\sigma}_{y}.
\end{align*}
Assuming weak coupling $g/\w_k\ll1$, the bosonic operators for modes with frequency $\omega_k>\Delta_0$ can be replaced by their expectation values
\begin{equation*}
\left\{
\begin{array}{l l}
&{\left\langle \cos\left(i\underset{\omega_k>\Delta_{0}}{\sum}\frac{g}{\w_{k}}(\hat{a}_{k}-\hat{a}_{k}^{\dagger})\right)\right\rangle }=\exp\left(-\frac{1}{2}\underset{\omega_k>\Delta_{0}}{\sum}\frac{g^{2}}{\w_{k}^{2}}\right),\\
&\left\langle \sin\left(i\underset{\omega_k>\Delta_{0}}{\sum}\frac{g}{\w_{k}}(\hat{a}_{k}-\hat{a}_{k}^{\dagger})\right)\right\rangle =0.
\end{array}
\right.
\end{equation*}
We finally obtain for the renormalized Hamiltonian $\hat H_1=\hat U_1 \hat H\hat U_1^\dagger$
\begin{equation}
\hat{H_1}=-\frac{\Delta_{1}}{2}\hat{\sigma}_{x}+\underset{k}{\sum}\omega_{k}\hat{a}_{k}^{\dagger}\hat{a}_{k}+\hat{\sigma}_{z}\underset{\omega_k\leq\Delta_{0}}{\sum}g(\hat{a}_{k}^{\dagger}+\hat{a}_{k})-\underset{\omega_k>\Delta_{0}}{\sum}\frac{g^{2}}{\omega_{k}},
\label{eq:H_1}
\end{equation}
where the constant term can be neglected and the reduced tunneling element reads
\begin{equation*}
\Delta_{1}=\Delta_0\exp\left(-2\underset{\omega_k>\Delta_{0}}{\sum}\frac{g^{2}}{\omega_{k}^{2}}\right).
\end{equation*}

Comparing eq.(\ref{eq:H_1}) to the original Hamiltonian we see that these transformations have reduced the tunneling strength and decreased the number of modes interacting with the qubit. Therefore another transformation can be applied to eliminate all modes with  $\Delta_0>\omega_k>\Delta_1$ and once again obtain a reduced tunneling element. The iterative transformation
\begin{equation*}
\hat{U}_{n}=\exp\left(-\hat{\sigma}_{z}\underset{\Delta_{n-2}>\omega>\Delta_{n-1}}{\sum}\frac{g}{\w_{k}}(\hat{a}_{k}-\hat{a}_{k}^{\dagger})\right),
\end{equation*}
with 
\begin{align*}
\Delta_{n}&=\Delta_{n-1}\exp\left(-2\underset{\Delta_{n-2}>\omega>\Delta_{n-1}}{\sum}\frac{g^{2}}{\omega_{k}^{2}}\right) \\
&= \Delta_0\exp\left(-2\underset{\omega>\Delta_{n-1}}{\sum}\frac{g^{2}}{\omega_{k}^{2}}\right),
\end{align*}
can be performed until $\Delta_n=\Delta_{n+1}$. The final renormalized tunneling element $\Delta_{\rm eff}$ must fulfil the self-consistency equation
\begin{equation*}
\Delta_{\rm eff}= \Delta_0\exp\left(-2\underset{\omega>\Delta_{\rm eff}}{\sum}\frac{g^{2}}{\omega_{k}^{2}}\right).
\end{equation*}

For a continuous spectrum, the renormalized tunneling element reads
\begin{equation}\label{eq:integralRenorm}
\Delta_{\text{eff}}=\Delta_{0}\exp\left(-2\int_{\Delta_{\text{eff}}}^{\infty}\frac{J(\w)}{\omega^{2}}d\w\right),
\end{equation}
where $ J(\w)=\underset{k}{\sum}g^{2}\delta(\w-\w_{k}) $ is the environmental spectral density. For the composite transmission line we use the DOM obtained in section \ref{sec:dom} and write $ J(\w)\approx g^{2}D(\w) $. Note that the density of modes did not take into account the presence of the qubit. A more detailed calculation has to incorporate its effects.

\subsection{Analytical Approach}
\begin{figure}[t]
\includegraphics[width=1\columnwidth]{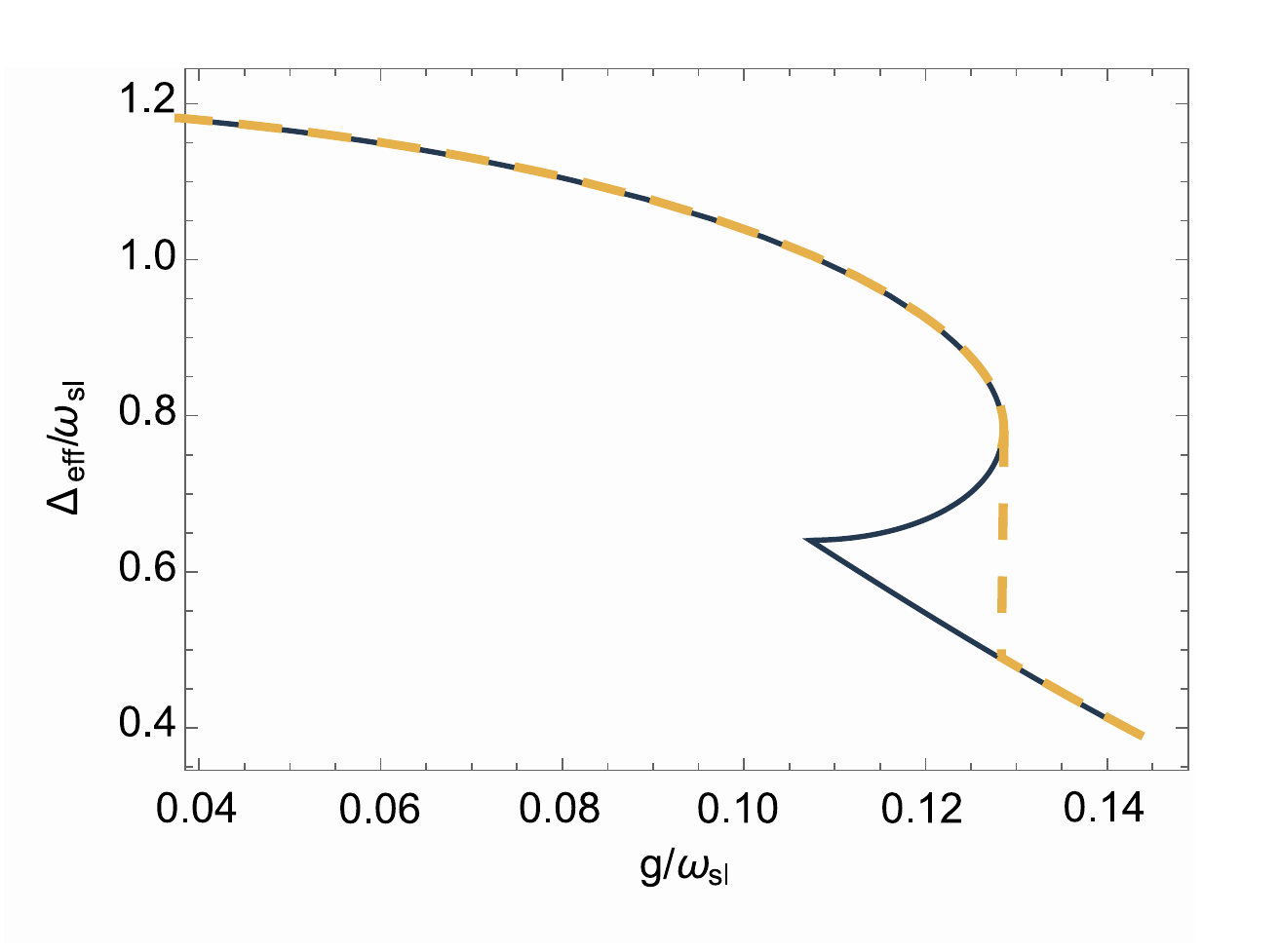}
\caption{(Color online) Typical solution to eq.(\ref{eq:integralRenorm}) on inverted axes for 20 superlattice cells with $\e=2$ and an initial energy splitting of the qubit of $\Delta_0=1.2\omega_{\rm sl}$. Blue (solid) line shows the coupling strength $g$ (horizontal axis) for a given renormalized tunelling element (vertical axis). Orange (dashed) line indicates the value to which $\Delta_{\rm eff}$ converges.
\label{fig:firsttrans}}
\end{figure}

The self-consistency equation, eq.\eqref{eq:integralRenorm}, can be inverted to give an expression for the coupling $g$ as a function of the effective tunneling, $\Delta_{\text{eff}}$. This requires an analytical solution to the integral. To obtain such a solution, we approximate the DOM obtained in sec.(\ref{sec:dom}) using a piecewise function describing the DOM in each of the bands,
\begin{equation}
D(\w)=\begin{cases}
{\displaystyle \frac{\alpha_{1}}{\sqrt{\w-\w_{\text{{\tiny{1-}}}}}\sqrt[\,4]{\w_{\text{{\tiny{1+}}}}-\w}},} & \w\in[\w_{\text{{\tiny{1-}}}},\w_{\text{{\tiny{1+}}}}],\\
{\displaystyle \frac{\alpha_{2}}{\sqrt{\w-\w_{2}}},} & \w\geq\w_{2},\\
{\displaystyle 0}, & \mathrm{elsewhere}.
\end{cases}
\end{equation}
Here $ \w_{\text{\tiny{1-}}}$ and $\w_{\text{\tiny{1+}}}$ are, respectively, the lower and upper band edges of the low frequency band and $\w_2$ is the band edge of the high frequency band edge. These are found by examining the limiting cases of the dispersion relation, eq.(\ref{eq:disp_sl}), and $ \alpha_1 $ and $ \alpha_2 $ are fitting parameters. The coupling constant for the first band can now be written as
\begin{equation}
g=\sqrt{\frac{\w_2^{3/2}\ln(\frac{\Delta_{\text{eff}}}{\Delta_{0}})}{-\pi\alpha_{2}}}
\end{equation}
and for the second band, 
\begin{equation*}
g= \sqrt{\frac{\w_2^{3/2}\ln(\frac{\Delta_{\text{eff}}}{\Delta_{0}})}{2\alpha_{2}\left(\frac{\w_{2}}{\Delta_{\text{eff}}}\sqrt{\frac{\Delta_{\text{eff}}}{\w_{2}}-1}+\arctan\left(\sqrt{\frac{\Delta_{\text{eff}}}{\w_{2}}-1}\right)-\frac{\pi}{2}\right)}}.
\end{equation*}
Figure (\ref{fig:firsttrans}) shows an instance of this solution. The dashed line shows the behavior of $\Delta_{\text{eff}}$ with increasing coupling, where we expect jumps when $g$ reaches a local maximum. 

\subsection{Phase Diagram} \label{subsec:PhaseDiag}
\begin{figure}[th]
	\includegraphics[width=1\columnwidth]{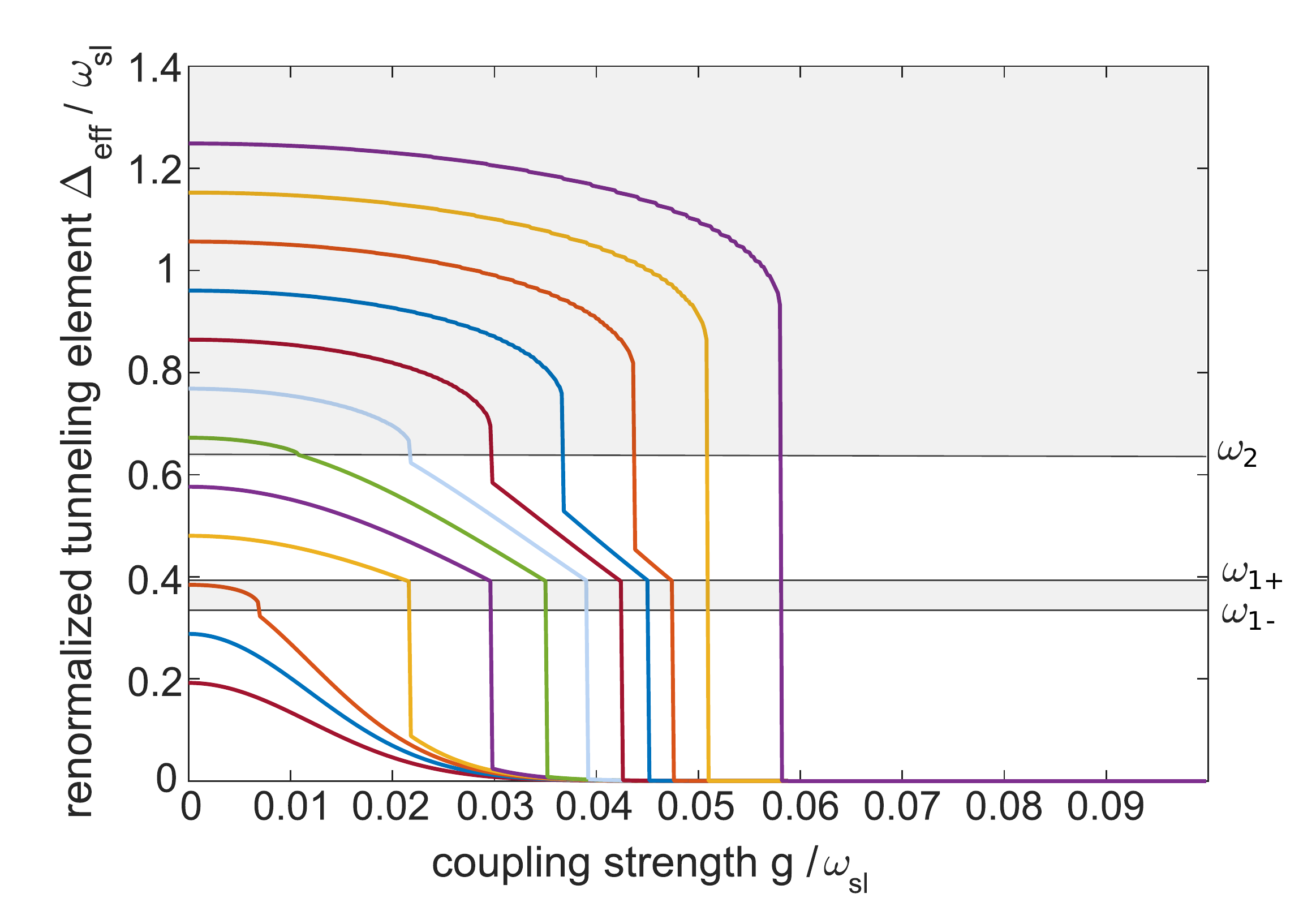}
	\caption{(Color online) Effective tunneling rate of the qubit as a function of the coupling constant (both in units of the superlattice resonance frequency $\omega_{\rm sl}$) for $\e=2$ and 200 superlattice cells. Each color represents a different initial bare tunnelling element, which coincides with the effective tunnelling element for zero coupling. Shaded regions represent the two energy bands. Localized, weakly localized and delocalized phases are clearly visible.
		\label{fig:jumps}}
\end{figure}

We now turn our attention to the phase diagram of our model. The results in this section were obtained using the iterative procedure of sec.(\ref{sec:ad_renorm}). From our previous discussion, we expect the existence of two phases, a localized one, with $\Delta_{\rm eff}=0$, and a delocalized one, with $\Delta_{\rm eff}\ne0$. Figure (\ref{fig:jumps}) shows the renormalized tunneling rate of the qubit as a function of the coupling constants, $g$. Different lines represent different bare tunneling rates, $\Delta_0$. As expected, for $g/\Delta_0$ sufficiently large, the system reaches the localized phase, whereas for $g/\Delta_0$ sufficiently small it remains in the delocalized phase, characterized by small corrections to the tunneling rate.

\begin{figure}[th]
\includegraphics[width=1\columnwidth]{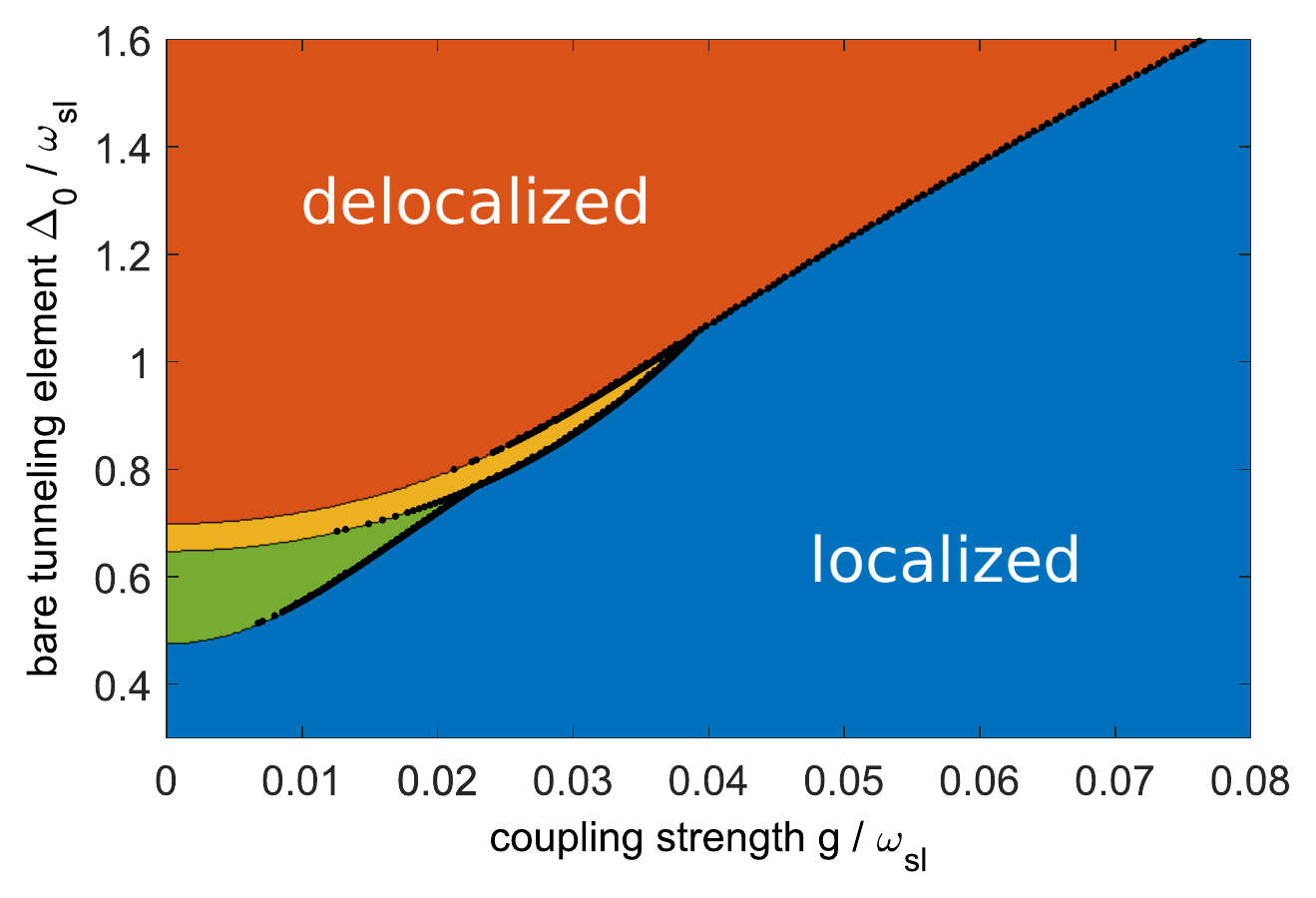}
\caption{(Color online) Phase diagram of the spin-boson model of a hybrid transmission line environment with $\e=1.1$ and 200 superlattice cells. Four distinct phases are shown. The colored background shows the energy region (separated by the band edges $\w_{\text{{\tiny{1-}}}}$, $\w_{\text{{\tiny{1+}}}}$ and $\w_{2}$)  $\Delta_{\rm eff}$ is lying in as a contour plot, whereas black dots are actual jumps found from numerics. For smaller coupling strengths jumps are less likely to be found numerically due to the decreased jump strength and the discrete nature of the superlattice.
\label{fig:phase_diagram}}
\end{figure}

Interestingly, two new phases appear for which $\Delta_{\rm eff}$ lies within the band gap or inside the first band. We term these, partially localized phases. These are characterized by strong renormalization of the tunneling rate, but not sufficient to reach the fully localized phase. Differently than the fully delocalized phase, for the phase lying in the band gap, phase transitions only occur as $\Delta_{\rm eff}$ reaches the band edge, entering the low frequency band. We note that because the superlattice used for this simulations is finite (with $n_{\rm sl}=200$ supercells), no genuinely localized phase is reached, as $\Delta_{\rm eff}\rightarrow0$ only in the infinite coupling limit. The qubit is thus in a quasi-localized phase. For an infinite superlattice the localized phase can be recovered. Figure (\ref{fig:phase_diagram}) shows the corresponding phase diagram, suggesting that jumps occur whenever the value of $\Delta_{\rm eff}$ crosses a band edge frequency.

Finally, we note that renormalization of the qubit energies originates from off-resonant degrees of freedom including those separated by a band gap. The band gap introduces a minimum detuning hence limiting the impact of each individual modes. The large density of modes at the band gap - the van Hove singularity - partially compensates for this suppression hence leading to a significant contribution even for qubit energies lying within the band gap. A more detailed model, where the frequency dependent qubit-mode coupling strength is taken into account is the object of future work.


\section{Conclusions}
Reservoir engineering is a cornerstone for many quantum technologies and systematic methods for implementing it are scarce. In this work we have thoroughly described how to apply superlattice, left-handed metamaterials as a means to manipulate a bosonic environment. The superlattice structure gives rise to a dual band spectrum, with the frequencies of the band edges controlled by the circuit parameters. This allows one to engineer spectra with a controlled number of modes in one band, while leaving the higher energy band isolated from the system of interest, i.e., the quantum system under investigation can be made to interact with a controlled number of modes. Moreover, these modes can have approximately the same coupling constant to the system, given the hybrid left-right-handed nature of the transmission line. The system proposed thus allows for specialized environment designs.

We have discussed one specific quantum simulation implementation, that of the phase diagram for the spin-boson model, where apart from a localized and a delocalized phase, we found two additional partially localized regions. The composite superlattice-right handed transmission line investigated here opens the possibility of exploring many different scenarios experimentally, like the strong coupling regime, for which the adiabatic renormalization technique discussed here should fail. Therefore experimental investigations would allow for the verification of the numerical methods and approximations.

The ideas presented in this work can straight-forwardly be adapted to other problems. Some possibilities include the quantum-classical transition and a systematic experimental study of the validity of some master equation approximations, such as local vs. global environments. Other applications can also be foreseen, such as the use of hybrid, superlattice transmission lines for filtering or to manipulate Purcell's effect on a qubit. Naturally, more general designs using left-handed transmission lines could be used to create more elaborate environmental spectra, suited for a number of different applications.

\section{Acknowledgements}
This work was supported by the Army Research Office under contract W911NF-14-1-0080 and the European Union through ScaleQIT. B.G.T. acknowledges support from Fapesc, , CNPq and the National Institute for Science and Technology - Quantum Information.

\appendix
\section{An alternative method to find the superlattice spectrum}
\label{app:EL_formalism}
Another approach to determine the dispersion relation is to use Euler-Lagrange formalism. The Lagrangian of the superlattice in terms of the magnetic flux $ \Phi $ reads
\begin{align*}
 \mathcal{L}=&\frac{1}{2}\sum_{n}\left[C_{\rm sl}(\dot{\Phi}_{n}-\dot{\Phi}_{n-1})^{2}+\epsilon C_{\rm sl}(\dot{\Phi}_{n}-\dot{\Phi}_{n+1})^{2}\right]\\
 &-\sum_n \left[\frac{1}{2\epsilon L_{\rm sl}} \Phi_{n}^{2}+\frac{1}{2L_{\rm sl}} \Phi_{n-1}^{2}\right].
\end{align*}
We use the Euler-Lagrange equation $ \frac{\mathrm d}{\mathrm{d}t}\frac{\partial  \mathcal{L}}{\partial \dot\Phi}-\frac{\partial  \mathcal{L}}{\partial \Phi}=0 $ to find the differential equations 
\begin{align*}
&C(\ddot{\Phi}_{2n}-\ddot{\Phi}_{2n-1})+\epsilon C (\ddot{\Phi}_{2n}-\ddot{\Phi}_{2n+1})+\frac{1}{\epsilon L}\Phi_{2n}=0,\\
&C(\ddot{\Phi}_{2n-1}-\ddot{\Phi}_{2n})+\epsilon C (\ddot{\Phi}_{2n-1}-\ddot{\Phi}_{2n-2})+\frac{1}{L}\Phi_{2n-1}=0.
\end{align*}
These equations can be combined, using $ \ddot{\Phi}=-\omega^2\Phi $ by assuming two independent wave equations in the two lattice elements and writing an equation for the even lattice cells 
\begin{align}
C^2\left(\frac{\Phi_{2j}+\epsilon \Phi_{2j-2}}{C+\epsilon C-\frac{1}{\omega^{2}L}}\right)+\epsilon C^2&\left(\frac{\Phi_{2j+2}+\epsilon \Phi_{2j}}{C+\epsilon C-\frac{1}{\omega^{2}L}}\right)\nonumber \\
&-(C+\epsilon C-\frac{1}{\omega^{2}\epsilon L})\Phi_{2j}=0.
\label{eq:evensites}
\end{align}
A plane wave ansatz valid only on even lattice sites
\begin{equation*}
\Phi_2n(t)=\Phi_{0}e^{i(kn\Delta z-\w t)}
\end{equation*}
can now be used to solve equation \ref{eq:evensites} for $\omega$ which yields the same result as equation \ref{eq:disp_sl}.



\bibliography{superlatticebib}
\bibliographystyle{apsrev4-1}

\end{document}